\documentclass[epsf,preprint,aps,12pt,showpacs,nofootinbib,tightenlines]{revtex4}
\usepackage{mathrsfs}
\usepackage{amsmath,rotating,calc,graphics,color}
\def\ps{p\hspace{-0.19cm}\slash}
\def\pars{\partial\hspace{-0.21cm}\slash}
\def\Ds{D\hspace{-0.25cm}\slash}
\begin{document}

\title{\Large{\bf TeV Scale Lee-Wick Fields out of \\Large Extra Dimensional Gravity } }
\author{Feng Wu$^{1}$}
\email[Electronic address: ]{fengwu@ncu.edu.cn}
\author{Ming Zhong$^{2,3}$}
\email[Electronic address: ]{zhongm@nudt.edu.cn}
\affiliation{${}^{1}$ Department of Physics, Nanchang University,
330031, China}
 \affiliation{${}^{2}$ Department of Physics, National
University of Defense Technology, Hunan 410073, China}
\affiliation{${}^{3}$ Kavli Institute for Theoretical Physics China,
CAS, Beijing 100190, China}
\begin{abstract}
We study the gravitational corrections to the Maxwell, Dirac and
Klein-Gorden theories in the large extra dimension model in which
the gravitons propagate in the (4+n)-dimensional bulk, while the
gauge and matter fields are confined to the four-dimensional world.
The corrections to the two-point Green's functions of the gauge and
matter fields from the exchanges of virtual Kaluza-Klein gravitons
are calculated in the gauge independent background field method. In
the framework of effective field theory, we show that the modified
one-loop renormalizable Lagrangian due to quantum gravitational
effects contains a TeV scale Lee-Wick partner of every gauge and
matter field as extra degrees of freedom in the theory. Thus the
large extra dimension model of gravity provides a natural mechanism
to the emergence of these exotic particles which were recently used
to construct an extension of the Standard Model.
\end{abstract}
 \maketitle 
\date{\today}
\section{Introduction}
 In the work of constructing a finite version of quantum electrodynamics \cite{Lee,Lee1}, Lee and Wick introduced a
physical massive vector field to play the role of the regulator in
Pauli-Villars regularization, and made the mass, charge and wave
function renormalizations be all finite. The Lee-Wick QED (LW QED)
is Poincar$\acute{e}$ invariant, gauge invariant and unitary.
Recently, Grinstein, O'Connell and Wise \cite{Grinstein} extended
this idea to the Standard Model (SM) and proposed that a higher
derivative term is added for each field in the SM. In this scenario
each propagator includes an extra degree of freedom corresponding to
a massive LW particle. The extended SM is shown to be free of
quadratic divergences. Therefore the mass of the Higgs particle is
stable and the hierarchy problem is solved. The Lee-Wick SM (LWSM)
does not provide any information on the masses of Lee-Wick
particles. However they are expected to have masses at TeV scale in
various phenomenological studies. Extensive discussions of the
phenomenology of the LWSM including its implications for LHC and
linear collider physics \cite{Krauss,Rizzo,Rizzo1}, neutrino physics
\cite{Espinosa}, the flavor changing neutral currents
\cite{Dulaney}, the electroweak precision constraints
\cite{Alvarez,Underwood} and another Lee-Wick extension of the SM
\cite{Carone} have been made. Some theoretical works on the
perturbative unitarity, the one-loop renormalization and the
causality problem of the LWSM have been discussed in
\cite{Grinstein1}, \cite{Grinstein2} and \cite{Grinstein3}
respectively. Chiral symmetry breaking and fermion mass generation
triggered by a higher derivative term was shown in \cite{gabrielli}.

We studied the Maxwell-Einstein theory in \cite{Wu}. We considered
the effect of gravity by expanding the metric around the flat
background spacetime and calculated the photon self-energy in the
framework of the gauge independent background field method. We
showed that the one-loop gravitational corrections induce a new
higher derivative term with mass dimension six, which is the term
needed in the LWSM. In a (3+1) dimensional renormalizable theory,
quantum corrections will generate possible UV divergences only to
the relevant and marginal operators whose mass dimensions are less
than five. This assures the predictiveness of the theory. General
relativity, the theory of gravitational interactions, on the other
hand, is not renormalizable after quantization \cite{DeWitt,tHooft}.
This is one of the reasons that general relativity was considered to
be incompatible with quantum mechanics. In the quantized version of
general relativity, one would not be able to reabsorb all the UV
divergences into the coupling constant in the original Lagrangian.
That is, new counterterms are needed at each order of perturbative
calculations when trying to renormalize the theory. Nevertheless, a
modern point of view is that a non-renormalizable theory might be
sensible and the reliable predictions could still be made from it
within the framework of effective field theories
\cite{Polchinski,Kaplan}. From the value of the only dimensional
coupling constant $G$, Newton's constant, in Hilbert-Einstein
Lagrangian, one can see that gravitational effects are tiny at
energies $E \ll M_{pl} \sim 10^{19} GeV/c^2$. It makes sense to
treat general relativity as a low energy effective field theory of
some unknown fundamental theory and consider its quantum effects
\cite{Donoghue}. The effects due to non-renormalizable terms are
suppressed by inverse powers of Planck scale $M_{pl}$, the mass
scale of new physics. In this sense, the higher derivative term
permitted by symmetry should be added to the Lagrangian at the
beginning so that the theory is one-loop renormalizable. The
modified Maxwell-Einstein theory contains a Lee-Wick vector field as
an extra degree of freedom. Since gravity does exist in Nature, it
provides a natural mechanism to the emergence of this exotic
particle. Different from the other scenarios, all the Lee-Wick
fields generated in this way have their masses of order $M_{pl}$,
and thus presently escape from any experimental measurement.

In this work we will investigate the gauge-matter-gravity system
where the gravitons propagate in (4+n) dimensional spacetime while
the gauge and matter fields live in the normal four dimensional
world. For the reason of phenomenological implications, we take the
compactification scale $1/R$ of the n extra dimensions to be as low
as $10^{-4} eV$. The Planck scale $M_{pl(4+n)}$ of the (4+n)
dimensional theory is taken to be of order $1~ TeV$, as the case in
the ADD model  \cite{Arkani-Hamed}. We will show, by explicit
calculations in the background field method, that up to one-loop
order, quantum corrections to each sector of the gauge, fermion and
scalar coming from the exchange of virtual Kaluza-Klein gravitons
will generate a new type of divergence which corresponds to a higher
derivative operator with mass dimension six. Summation of the
Kaluza-Klein towers greatly improves the coefficients of these
operators up to $\sim 1~ TeV^{-2}$. Therefore, one needs to modify
the Lagrangian of the system and include the higher derivative
operators at the beginning to absorb the divergent quantum
corrections. This modification is natural and consistent with the
framework of effective field theories. The modified theory now
contains a Lee-Wick partner whose mass is at $TeV$ scale for every
gauge, fermion and scalar particle. On the one hand, these Lee-Wick
particles are the necessary components of the extension of the SM in
\cite{Grinstein} and turn out to be interesting and significant in
describing the physics at $TeV$ scale. On the other hand, to our
knowledge, they were ignored in the relevant phenomenological
studies of the ADD model in literatures.

The rest of the paper is organized into three parts. In Sec.
\ref{sec2}, we describe the Maxwell, Dirac and Klein-Gorden theories
in the context of the ADD gravity. We compactify the gravity on an
n-dimensional torus $T^n$ and perform a Kaluza-Klein decomposition.
In Sec. \ref{sec3}, we compute the one-loop gravitational
corrections to the gauge, fermion and scalar self-energy in the
framework of the background field method and discuss the
implications of the higher derivative operators. The similar
calculations on the various self-energy corrections have been made
in the conventional gauge dependent method \cite{Han,Han1}, where a
cutoff procedure has been used in the loop momentum integrations and
the summations of the Kaluza-Klein states. In this work we use the
dimensional regularization so that one can easily track out the
momenta of the background fields. We present our discussions and
conclusions in the final section.

\section{The Formalism and Kaluza-Klein Decomposition of the Gravity}
\label{sec2}
   Our starting point is the Hilbert-Einstein, Maxwell, Dirac and Klein-Gorden theory in which the gravity propagates in (4+n)-dimensional bulk while the gauge
and matter fields live in the four-dimensional spacetime. The extra
dimensions are compactified on a torus $T^n$. The action of the
theory has the form
\begin{equation}
S= - \int d^4x d^ny ({\cal L}_{HE}+{\cal L}_{M}+{\cal L}_{D}+{\cal
L}_{KG}), \label{action}
\end{equation}
where the Lagrangians for gravity and gravity-gauge-matter couplings are
\begin{eqnarray}
{\cal L}_{HE}&=&{1\over \hat{\kappa}^2} \sqrt{ (-1)^{3+n}|\hat{g}^{(4+n)}|}\hat{R}, \nonumber\\
{\cal L}_{M}&=&{1\over4}\sqrt{ - |\hat{g}^{(4)}|} \hat{g}^{\mu \lambda} \hat{g}^{\nu \rho} F_{\mu \nu} F_{\lambda \rho}\delta^{(n)}(y), \nonumber\\
{\cal L}_{D}&=&-\sqrt{ - |\hat{g}^{(4)}|}\bar{\psi}[i\hat{e}^{\mu}_{\,\alpha}\gamma^{\alpha}(D_{\mu}+\frac{1}{2}\hat{\omega}_{\mu
ab}\sigma^{ab})]\psi\delta^{(n)}(y),\nonumber\\
 {\cal L}_{KG}&=&-{1\over2}\sqrt{ -
|\hat{g}^{(4)}|}(\hat{g}^{\mu
\nu}\partial_{\mu}\varphi\partial_{\nu}\varphi-m^2_s\varphi^2)\delta^{(n)}(y),
\end{eqnarray} with $\hat{\kappa}^2 \equiv 16 \pi \hat{G}_N\sim
\frac{1}{M_{pl(4+n)}^{2+n}}$. The hatted symbols $\hat{R}$,
$\hat{g}^{\hat{\mu} \hat{\nu}}$ and $\hat{G}$ are the Ricci scalar,
metric tensor and the Newton constant in (4+n) bulk. The index
$\hat{\mu}$ extends over the full (4+n) dimensions and $\mu$ over
the (3+1) dimensions. The vierbein $\hat{e}_{\mu}^{\,\nu}$ is
defined by
$\hat{e}_{\mu}^{\,\alpha}\hat{e}_{\nu}^{\,\beta}\eta_{\alpha\beta}=\hat{g}_{\mu\nu}$.
The spin connection $\hat{\omega}_{\mu ab}$ can be solved in terms
of the vierbein and $\sigma^{ab}={1\over 4}[\gamma^a,\gamma^b]$. The
derivative $D_{\mu}$ is internal gauge symmetry covariant. The
fermion is massless because we are interested in the case of high
energies when electroweak symmetry is unbroken. The scalar field is
taken to be real and thus it does not carry any nontrivial charges.
We ignore the cosmological constant term since it is irrelevant to
our discussions. The action (1) is invariant under general
coordinate and $U(1)$ gauge transformations. For simplicity, we
assume that the compactification scales of the extra n-dimensional
spaces $y_i$ are all roughly equal to $R$.

In the following we use the background field method \cite{Abbott}
and choose the background spacetime to be Minkowski space. The basic idea of
the method is to expand the fields appearing in the classical action
to the background fields and the quantum fields. The quantum fields
are the variables of integration in the functional integral. A gauge choice is made in such a way that it breaks the gauge invariance of the
quantum gauge field, but retains gauge invariance in terms of the
background gauge field. Therefore, one is able to quantize a gauge
field theory without losing the explicit gauge invariance. The
Green's functions obey the naive Ward identities of gauge invariance
and even the unphysical quantities like divergent counterterms take
a gauge invariant form, which makes the following discussion
unambiguous. The background field method is used extensively in
gravity. The first relevant papers calculating the one-loop quantum
gravitational effects on the gauge and matter fields are
\cite{tHooft,Deser}.

Graviton field $\hat{g}_{\hat{\mu}\hat{\nu}}(x,y)$, gauge field
$A(x)$ and matter fields $\psi(x)$ and $\varphi(x)$ can be written
as sums of background fields $(\hat{\eta}_{\hat{\mu}\hat{\nu}},
\bar{A}(x), \Psi(x), \Phi(x))$ and quantum fluctuations $(
\hat{h}_{\hat{\mu}\hat{\nu}}, a(x), \tilde{\psi}(x),
\tilde{\varphi}(x))$:
\begin{eqnarray}
&& \hat{g}_{\hat{\mu}\hat{\nu}}(x,y)=\hat{\eta}_{\hat{\mu}\hat{\nu}}
+ \hat{\kappa} \hat{h}_{\hat{\mu}\hat{\nu}}(x,y) , \;\;\;\;  A(x) =
\bar{A}(x) +
a(x) , \nonumber\\
&&\psi(x)=\Psi(x)+\tilde{\psi}(x), \;\;\;\;\;\;\;\;\;\;\;\;\;\;\;\;
\varphi(x)=\Phi(x)+\tilde{\varphi}(x). \label{fields}
\end{eqnarray}
Here $\hat{\eta}_{\hat{\mu}\hat{\nu}}$ is the Minkowski metric of the bulk.

Now we consider the Hilbert-Einstein term in action (1). Expanding
the gravity field, one can get the linearized Fierz-Pauli Lagrangian
\begin{eqnarray} \frac{-\sqrt{ (-1)^{3+n}|\hat{g}^{(4+n)}|}
\hat{R}(x,y)}{\hat{\kappa}^2}&=&\frac{1}{4}(\partial^{\hat{\mu}}\hat{h}\partial_{\hat{\mu}}\hat{h}-\partial^{\hat{\mu}}\hat{h}^{\hat{\nu}\hat{\rho}}\partial_{\hat{\mu}}\hat{h}_{\hat{\nu}\hat{\rho}}
+2\partial_{\hat{\nu}}\hat{h}^{\hat{\nu}\hat{\mu}}\partial^{\hat{\rho}}\hat{h}_{\hat{\rho}\hat{\mu}}-2
\partial_{\hat{\nu}}\hat{h}^{\hat{\nu}\hat{\mu}}\partial_{\hat{\mu}}\hat{h})+{\cal O}(\hat{\kappa}), \nonumber\\
&& \label{F-P}
\end{eqnarray}
where $\hat{h}\equiv \hat{h}^{\hat{\mu}}_{\;\hat{\mu}}$. It is the
kinematic of the graviton, and we have ignored the self-interaction
terms of gravitons since they are irrelevant to our present
discussions.

To perform the Kaluza-Klein reduction of Eq. (\ref{F-P}) to
four-dimensional spacetime, we parameterize the field
$\hat{h}_{\hat{\mu}\hat{\nu}}$ as
\begin{equation}
\hat{h}_{\hat{\mu}\hat{\nu}}=V^{-{1\over2}}_n\left(
  \begin{array}{cc}
    h_{\mu\nu}+\eta_{\mu\nu}\phi & A_{\mu j} \\
    A_{i\nu} & 2\phi_{ij} \\
  \end{array}
\right),
 \end{equation}
where $V_n=R^n$ is the volume of the $n$-dimensional compactified
torus $T^n$, $\phi=\sum_i\phi_{ii}$, the subscript $\mu,\nu=0,1,2,3$
and $i,j=4,5,...,3+n$. The fields $h_{\mu\nu}$, $A_{\mu i}$ and
$\phi_{ij}$ are Lorentz tensor, vector and scalar respectively. They
have the following mode expansions:
\begin{eqnarray}
h_{\mu\nu}(x,y)&=&\sum_{\vec{n}}h_{\mu\nu}^{\vec{n}}(x)exp(i\frac{2\pi\vec{n}\cdot\vec{y}}{R}),\\
A_{\mu i}(x,y)&=&\sum_{\vec{n}}A_{\mu i}^{\vec{n}}(x)exp(i\frac{2\pi\vec{n}\cdot\vec{y}}{R}),\\
\phi_{ij}(x,y)&=&\sum_{\vec{n}}\phi_{ij}^{\vec{n}}(x)exp(i\frac{2\pi\vec{n}\cdot\vec{y}}{R}),
\end{eqnarray}
with $\vec{n}={(n_1,n_2,...,n_n)}$. After a straightforward calculation, we find that Eq. (\ref{F-P}) reduces to the expression in terms of the massive
Kaluza-Klein modes
\begin{eqnarray}
&&-\frac{1}{4}\sum_{\vec{n}}(2\partial^{\mu}A^{\vec{n},i\nu}\partial_{\mu}A^{-\vec{n}}_{\;i\nu}-2\partial_{\mu}A^{\vec{n},i\mu}\partial_{\nu}A^{-\vec{n},i\nu}-2m_{\vec{n}}^2A^{\vec{n},i\mu}
A^{-\vec{n}}_{\;i\mu}-2m_{n_i}m_{n^j}A^{\vec{n},i\mu}
A^{-\vec{n}}_{\;j\mu}\nonumber\\
&&\hspace{-10pt}+\partial^{\mu}h^{\vec{n},\nu\rho}\partial_{\mu}h^{-\vec{n}}_{\;\nu\rho}-
\partial^{\mu}h^{\vec{n}}\partial_{\mu}h^{-\vec{n}}-2\partial_{\nu}h^{\vec{n},\nu\mu}\partial^{\rho}h^{-\vec{n}}_{\;\rho\mu}
+2\partial_{\mu}h^{\vec{n}}\partial_{\nu}h^{-\vec{n},\mu\nu}-m_{\vec{n}}^2h^{\vec{n},\mu\nu}h^{-\vec{n}}_{\;\mu\nu}+m_{\vec{n}}^2h^{\vec{n}}h^{-\vec{n}}\nonumber\\
&&\hspace{-10pt}+2\partial^{\mu}\phi^{\vec{n}}\partial_{\mu}\phi^{-\vec{n}}+4\partial^{\mu}\phi^{\vec{n},ij}\partial_{\mu}\phi^{-\vec{n}}_{ij}
-4m_{\vec{n}}^2\phi^{\vec{n},ij}\phi^{-\vec{n}}_{ij}-8m_{n_i}m_{n^j}\phi^{\vec{n},ik}\phi^{-\vec{n}}_{jk}+8m_{n_i}m_{n_j}\phi^{\vec{n}}\phi^{-\vec{n},ij}\nonumber\\
&&\hspace{-10pt}+2m_{\vec{n}}^2h^{\vec{n}}\phi^{-\vec{n}}+4m_{n_i}m_{n_j}h^{\vec{n}}\phi^{-\vec{n},ij}
+i4m_{n^i}\partial_{\mu}A^{\vec{n},j\mu}\phi^{-\vec{n}}_{ij}+i4m_{n_i}\phi^{\vec{n}}\partial_{\mu}A^{-\vec{n},i\mu}\nonumber\\
&&\hspace{-10pt}+i4m_{n^i}\partial_{\mu}h^{\vec{n},\mu\nu}A^{-\vec{n}}_{i\nu}+i2
m_{n_i}h^{\vec{n}}\partial_{\mu}A^{-\vec{n},i\mu}-i2m_{n_i}\partial_{\mu}h^{\vec{n}}A^{-\vec{n},i\mu}),\label{mixing}
\end{eqnarray}
where $m_{\vec{n}}^2\equiv
-m_{n_i}m_{n^i}=-\frac{4\pi^2n_in^i}{R^2}=\frac{4\pi^2n_in_i}{R^2}$
since we have used the flat spacetime metric tensor $
diag(\hat{\eta})=(1,-1,...,-1)$. In deriving Eq. (\ref{mixing}), we
have use the relation between the four-dimensional and the
(4+n)-dimensional Newton's constant $\kappa^2R^n=\hat{\kappa}^2$.
The last two lines contain the mixing terms $h\phi$, $A\phi$ and
$hA$. All of them result from the last two terms in Eq. (\ref{F-P}).
It is more convenient to cancel such terms in practical
calculations. For this purpose, we add a special de Donder gauge
fixing term
\begin{equation}-{1 \over 2}(
\partial_{\hat{\rho}} \hat{h}^{\hat{\rho} \hat{\mu}} \partial^{\hat{\sigma}} \hat{h}_{\hat{\sigma} \hat{\mu}}- \partial_{\hat{\rho}} \hat{h}^{\hat{\rho}
\hat{\mu}}
\partial_{\hat{\mu}} \hat{h}+\frac{1}{4}\partial_{\hat{\mu}} \hat{h}\partial^{\hat{\mu}} \hat{h})\end{equation}
to the Fierz-Pauli Lagrangian (\ref{F-P}), rather than redefine the fields as in \cite{Han}. The first two terms will cancel the last two terms in Eq.
(\ref{F-P}) so that the mixing terms in Eq. (\ref{mixing}) do not appear again. The Lagrangian of the gravity can then be written as a simple form
\begin{eqnarray}
{\cal
L}_{HE}&=-&\frac{1}{4}\sum_{\vec{n}}(\partial^{\mu}h^{\vec{n},\nu\rho}\partial_{\mu}h^{-\vec{n}}_{\;\nu\rho}-
\frac{1}{2}\partial^{\mu}h^{\vec{n}}\partial_{\mu}h^{-\vec{n}}-m_{\vec{n}}^2h^{\vec{n},\nu\rho}h^{-\vec{n}}_{\;\nu\rho}+\frac{1}{2}m_{\vec{n}}^2h^{\vec{n}}h^{-\vec{n}}
\nonumber\\
&&+2\partial^{\mu}A^{\vec{n},i\nu}\partial_{\mu}A^{-\vec{n}}_{\;i\nu}-2m_{\vec{n}}^2A^{\vec{n},i\nu}
A^{-\vec{n}}_{\;i\nu}\nonumber\\
&&+4\partial^{\mu}\phi^{\vec{n},ij}\partial_{\mu}\phi^{-\vec{n}}_{ij}+2\partial^{\mu}\phi^{\vec{n}}\partial_{\mu}\phi^{-\vec{n}}
-4m_{\vec{n}}^2\phi^{\vec{n},ij}\phi^{-\vec{n}}_{ij}-2m_{\vec{n}}^2\phi^{\vec{n}}\phi^{-\vec{n}})+{\cal
O}(\kappa).
\end{eqnarray}
In this gauge, the propagators for the massive Kaluza-Klein states $h^{\vec{n}}_{\mu\nu}$, $A^{\vec{n}}_{i\mu}$ and $\phi^{\vec{n}}_{ij}$ are
\begin{eqnarray}
&&\triangle^h_{\vec{n}\mu\nu,\vec{m}\rho\sigma}(k)=-i\frac{\delta_{\vec{n},-\vec{m}}(\eta_{\mu\rho}\eta_{\nu\sigma}+\eta_{\mu\sigma}\eta_{\nu\rho}-\eta_{\mu\nu}\eta_{\rho\sigma})}
{k^2-m_{\vec{n}}^2+i\epsilon}\\
&&\triangle^A_{\vec{n}i\mu,\vec{m}j\nu}(k)=-i\frac{\delta_{\vec{n},-\vec{m}}\delta_{ij}\eta_{\mu\nu}}{k^2-m_{\vec{n}}^2+i\epsilon}\\
&&\triangle^\phi_{\vec{n}ij,\vec{m}kl}(k)=-i\frac{\delta_{\vec{n},-\vec{m}}[\frac{1}{4}(\delta_{ik}\delta_{jl}+\delta_{il}\delta_{jk})-\frac{1}{4+2n}\delta_{ij}\delta_{kl}]}
{k^2-m_{\vec{n}}^2+i\epsilon}.
\end{eqnarray}
Obviously the spin-2 state $h^{\vec{n}}_{\mu\nu}$ is not a physical state. This can be seen explicitly from the numerator of its propagator. The physical massive spin-2 state with right polarization tensor is  constructed from $h^{\vec{n}}_{\mu\nu}$, $A^{\vec{n}}_{i\mu}$ and $\phi^{\vec{n}}_{ij}$. It's explicit form is shown in \cite{Han}. However, the rearrangement of the physical spin-2, $(n-1)$ spin-1, and $n(n-1)/2$ spin-0 states into $h^{\vec{n}}_{\mu\nu}$, $A^{\vec{n}}_{i\mu}$ and $\phi^{\vec{n}}_{ij}$ states shown above greatly simplifies our calculations.

Note that we do not include ghost parts in the Lagrangian since they do not contribute to the one-loop order.

\section{Lee-Wick Particles from Gravitational Corrections}
\label{sec3} Now we turn to the interactions of the gravity with the
gauge and matter fields. What we are concerned with are the higher
derivative operators of mass dimension six from the one-loop
gravitational corrections to the two-point Green's functions for the gauge and matter wave operators. The coefficients of the quantum induced higher
derivative operators are divergent. In terms of the effective
theory, one should include each higher derivative operator with an
arbitrary parameter $a_i(i=1,2,3)$ in the Lagrangian, then introduce
counterterms to cancel the divergences and renormalize the
parameters $a_i$. The tadpole diagrams, though they do not vanish
because of the appearing of the massive Kaluza-Klein states in the
loops, are irrelevant to the higher derivative operators. Thus in
what follows, we only need to calculate the rainbow diagrams and
write down the divergent terms explicitly. We will present a rather
detailed description on the gauge field in Sec. \ref{sub1}, and
leave out the similar ones on the fermion and scalar field sectors.
We would like to emphasize that since gravitons do not carry gauge
charges, the results of the one-loop corrections to the two-point
Green's functions can be applied directly to the non-abelian case.

\subsection{Lee-Wick Gauge Bosons}
\label{sub1}
 The second term in action (1) specifies the interaction of the gauge field with the graviton. Adding the Lorentz gauge fixing
term ${-1\over 2 \xi} (
\partial_{\mu} a_{\nu} )^2$ for the photon field to the action, we find the propagator for the photon $a_\mu$
\begin{equation}
\triangle^a_{\mu\nu}(k)={-i \over k^2 + i \epsilon} [ \eta_{\mu\nu} -(1- \xi) { k_{\mu} k_{\nu} \over k^2}]. \end{equation}

Up to order $\kappa^2$, the relevant interaction terms contain
($h^{\vec{n}}a\bar{A}$), ($h^{\vec{n}}h^{\vec{m}}\bar{A}\bar{A}$)
and ($h^{\vec{n}}\phi^{\vec{m}}\bar{A}\bar{A}$) vertexes, from which
only the first one is of interest in our present consideration since
it is the only vertex to compose rainbow diagram shown in Fig. 1.
\begin{eqnarray}
{\cal
L}_{M}^{int}&\sim&\kappa\sum_{\vec{n}}[h^{\vec{n}}_{\mu\nu}\partial_{\lambda}
\bar{A}_{\rho}\partial_{\tau}a_{\sigma}(\eta^{\lambda\tau}\eta^{\nu\sigma}
\eta^{\mu\rho}+\eta^{\mu\lambda}\eta^{\rho\sigma}
\eta^{\nu\tau}+\frac{1}{2}\eta^{\mu\nu}\eta^{\rho\tau}
\eta^{\lambda\sigma}\nonumber\\
&&\hspace{3.4cm}-\eta^{\lambda\sigma}\eta^{\mu\rho}\eta^{\nu\tau}-\eta^{\mu\lambda}\eta^{\rho\tau}
\eta^{\nu\sigma}-\frac{1}{2}\eta^{\mu\nu}\eta^{\lambda\tau}\eta^{\rho\sigma})]+{\cal O}(\kappa^2).
\end{eqnarray}

\begin{figure}[t]
\begin{center}
\includegraphics[width=13cm,clip=true,keepaspectratio=true]{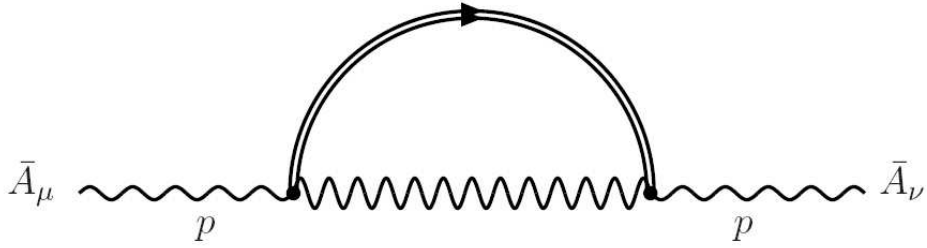}
\caption{\small The rainbow diagram generating the higher derivative
term of gauge field. The internal wavy and double lines represent
$a_{\mu}$ and $h^{\vec{n}}_{\mu\nu}$ fields respectively. The
external wavy lines are background photon fields.}
\end{center}\label{diagram1}
\end{figure}

The diagram is of order $\kappa^2$ of the gravitational coupling. We
evaluate the loop integral in dimensional regularization (DR) scheme
to extract the singularity, rather than introduce a hard cutoff as
in \cite{Han,Han1}. By a straightforward calculation, it can be
shown that the correction obtained in the framework of background
field method is independent of the gauge parameter $\xi$, as it
should be. The result of the diagram is
\begin{eqnarray}
\Pi^R_{\mu\nu}(p^2)&=&i\frac{\kappa^2}{24\pi^2}{1 \over \epsilon} p^2 (p^2 \eta_{\mu\nu}- p_{\mu} p_{\nu})\sum_{\vec{n}} \textbf{1} \nonumber\\
 &&+i\frac{3\kappa^2}{8\pi^2}{1 \over \epsilon} (p^2 \eta_{\mu\nu}- p_{\mu} p_{\nu})\sum_{\vec{n}}m^2_{\vec{n}}+ \; [finite\;\; part].
\label{resultg}
\end{eqnarray}
The contributions of the extra dimensions are embodied in the
summation over the Kaluza-Klein states in a tower. In the limit of
four-dimensional spacetime, the second term is vanished because the
graviton is massless. The summation can be written as an integration
in term of the mass $m^2_{\vec{n}}$ when when the Kaluza-Klein
states are nearly degenerate \cite{Han}
\begin{equation}
\sum_{\vec{n}}f(m_{\vec{n}})=\int^{\Lambda^2}_0
dm^2_{\vec{n}}\rho(m_{\vec{n}})f(m_{\vec{n}}),
\end{equation}
where
\begin{equation}
\rho(m_{\vec{n}})=\frac{R^n m^{n-2}_{\vec{n}}}{(4\pi)^{n/2}\Gamma(n/2)}
\end{equation}
is the Kaluza-Klein state density. As is well-known, using DR
scheme, which is a mass-independent scheme, heavy states do not
decouple. Thus we have introduced an explicit cutoff $\Lambda$ to
regularize the mass integration. That is, we include only a finite
number of low-lying Kaluza-Klein states and assume all other states
decouple from the low energy physics we are interested. This cutoff
does not break any gauge symmetry and our calculation is gauge
independent. The nearly degenerate condition is satisfied when the
energy scale $R^{-1}$ characterized the Kaluza-Klein excitations is
much less than the physical scale $\Lambda$. It is indeed the case
in large extra dimension model. In general, we have $\Lambda\leq
M_{pl(4+n)}$, since the effective theory is only expected to be
valid below the fundamental scale $M_{pl(4+n)}$. Various of
phenomenological studies suggested that the cutoff should be
$\Lambda\sim M_{pl(4+n)}\sim 1~ TeV$\cite{Han,Han1,Giudice}. Our
final result for the rainbow diagram is
\begin{eqnarray}
\Pi^R_{\mu\nu}(p^2)&=&i\frac{\hat{\kappa}^2}{12\pi^2}[\frac{\Lambda^n}{(4\pi)^{n/2}\Gamma(n/2)n}]{1 \over \epsilon} p^2 (p^2 \eta_{\mu\nu}- p_{\mu} p_{\nu})\nonumber\\
 &&+i\frac{3\hat{\kappa}^2}{4\pi^2}[\frac{\Lambda^{n+2}}{(4\pi)^{n/2}\Gamma(n/2)(n+2)}]{1 \over \epsilon} (p^2 \eta_{\mu\nu}- p_{\mu} p_{\nu})\nonumber\\
 &&+ \; [finite\;\; part].
\label{resultg1}
\end{eqnarray}
The first term of Eq. (\ref{resultg1}) shows that different from
renormalizable theories, the correction due to gravitons generates a
new type of divergence which cannot be absorbed in the
Maxwell-Einstein action. In flat spacetime, the term
$ip^2(p_{\mu}p_{\nu} - p^2 \eta_{\mu\nu})$ in the truncated
photon-photon correlation function corresponds to the dimension six
operator $-{1\over 2}
\partial _{\mu}\bar{F}^{\mu\nu}
\partial^{\rho}\bar{F}_{\rho\nu}$. This is the leading higher derivative term allowed by symmetries.
Without renormalizability as an axiom, it should be included
in the Lagrangian. If one unnaturally neglects it, the theory will
lack its predictiveness at one-loop order. The second term which
vanishes in the 4-dimensional theory results from the summation of
the Kaluza-Klein states. It contributes to the one-loop
$\beta$-function of the gauge coupling and leads to the power law
running of the gauge coupling.

Note that the one-loop correction to the gauge couplings from a
tower of gauge boson Kaluza-Klein states is of order $\Lambda^{n}$
\cite{Dienes,Dienes1}. Here we show explicitly that the correction
coming from graviton Kaluza-Klein states is of order
$\Lambda^{n+2}$. The effects on the unification of gauge coupling
constants caused by this term is itself an interesting subject and
will be reported elsewhere \cite{wuzhong}.

When we take $n=0$ in Eq. (\ref{resultg1}), we will come to the case
of gravity in four-dimensional spacetime that we have discussed in
\cite{Wu}. But since we start from a different definitions of the
action (\ref{action}) and $\kappa$ in this work, the result here
differs by a factor of ${1\over 8}$ from that in \cite{Wu}.

Now let us consider the modified Maxwell theory in the curved
spacetime with the required higher derivative term. The action has
the form:
\begin{eqnarray}
- \int d^4xd^ny [\sqrt{ - |\hat{g}^{(4)}|}({1\over4} \hat{g}^{\mu \lambda} \hat{g}^{\nu \rho} F_{\mu \nu} F_{\lambda \rho}-{a_{1} \over 2M^2}\hat{g}^{\mu
\rho}\hat{g}^{\nu \lambda}\hat{g}^{\sigma \tau} {\cal D}_{\mu} F_{\rho\sigma} {\cal D}_{\nu} F_{\lambda\tau} )\delta^{(n)}(y)] \label{modlag}
\end{eqnarray}
where $a_{1}$ is a dimensionless parameter and will be renormalized
by introducing a counterterm to cancel the divergence of the first
term in Eq. (\ref{resultg1}) at certain renormalization condition.
The operator ${\cal D}_{\mu}$ is the spacetime covariant derivative.
For convenience, we have defined a dimension one parameter
$M\equiv\sqrt{1/\hat{\kappa}^2\Lambda^n}\sim 1~ TeV$. Based on the
previous discussion the origin of the last term is clear. The
existence of the gravity naturally provides a mechanism to generate
this non-renormalizable term. With this term the gauge sector in the
theory is one-loop renormalizable.

Let us consider again the four-dimensional background spacetime to
be a Minkowski one and focus our discussion on the gauge sector. The
part of the action we are interested is the quadratic terms of the
gauge field
\begin{equation}
- \int d^4x ({1\over4} F_{\mu \nu} F^{\mu \nu}-{a_{1} \over 2M^2}
\partial_{\mu} F^{\mu\nu} \partial^{\rho} F_{\rho\nu} ) ,
\label{propagator}
\end{equation}
from which we can find the propagator in the background spacetime:
\begin{equation}
{-i \over p^2 - {a_{1} \over M^{2}}p^4+i \epsilon} [ \eta_{\mu\nu} -
{p_{\mu}p_{\nu} \over q^2}+ \xi (1-a_{1} {p^2 \over M^2} ){p_{\mu}
p_{\nu} \over p^2} ], \label{propagator1}
\end{equation}
where we have use the same Lorentz gauge fixing term as before. The
propagator contains two poles: One corresponds to the massless
photon and the other one at $p^2={M^2 \over a_{1}}$ corresponds to
the Lee-Wick particle with mass ${M\over \sqrt{a_1}}$ for positive
$a_{1}$.

Remember the electric field $E^{i}=F^{i0}$ and magnetic fields $B^{i}=-\epsilon^{ijk} F_{jk}$. The ``Maxwell's equations" derived from Eq. (\ref{propagator}) reads
\begin{eqnarray}
\vec{\nabla}\cdot \vec{B}=0,\\
{\partial \vec{B} \over \partial t} + \vec{\nabla}\times \vec{E} =0,\\
\vec{\nabla}\cdot \vec{D}=0,\\
{\partial \vec{D} \over \partial t} - \vec{\nabla}\times \vec{H} =0,
\end{eqnarray}
where
\begin{eqnarray}
\vec{D} \equiv \left(1+ {a_{1} \over M^2}\left({\partial^2 \over \partial t^2} - \vec{\nabla}^2\right)\right) \vec{E},\\
\vec{H} \equiv \left(1+ {a_{1} \over M^2}\left({\partial^2 \over \partial t^2} - \vec{\nabla}^2\right)\right) \vec{B}.
\end{eqnarray}
The first set comes from the gauge invariance of the system. The second set is the equations of motion derived from Eq. (\ref{propagator}). One can easily see that the solution $\vec{D}=\vec{E}$ and $\vec{H}=\vec{B}$ corresponds to the well-known massless photon. The other independent solution, $\vec{D}=0$ and $\vec{H}=0$, implies
\begin{eqnarray}
({\partial^2 \over \partial t^2} - \vec{\nabla}^2 +{M^2 \over a_{1}} )\vec{E} =0, \\
({\partial^2 \over \partial t^2} - \vec{\nabla}^2 +{M^2 \over a_{1}} ) \vec{B}=0.
\end{eqnarray}
This corresponds to the Lee-Wick particle with mass ${M\over \sqrt{a_1}}$.
\subsection{Lee-Wick Fermions}
We now switch to the third term in the action (\ref{action})
which contains the interaction terms of the fermion field with the gravitons. Since only the
rainbow diagrams plotted in Fig. 2 can contribute to the higher
derivative operator, we are interested in the following interaction
terms exclusively
\begin{eqnarray}
&&{\kappa \over 2}\sum_{\vec{n}}\{
\bar{\tilde{\psi}}[i\gamma^{\mu}(\partial_bh^{\vec{n}}_{a\mu}+\eta_{a\mu}\partial_b\phi^{\vec{n}})\sigma^{ab}-i(h^{\vec{n},\mu}_{\alpha}+\delta^{\mu}_{\,
\alpha}\phi^{\vec{n}})\gamma^{\alpha}\partial_{\mu} \nonumber\\
&&\quad\quad\quad\quad +(h^{\vec{n}}+4\phi^{\vec{n}})i\gamma^{\mu}\partial_{\mu}]\Psi \nonumber\\
&&\quad\quad
+\bar{\Psi}[i\gamma^{\mu}(\partial_bh^{\vec{n}}_{a\mu}+\eta_{a\mu}\partial_b\phi^{\vec{n}})\sigma^{ab}-i(h^{\vec{n},\mu}_{\alpha}+\delta^{\mu}_{\,
\alpha}\phi^{\vec{n}})\gamma^{\alpha}\partial_{\mu} \nonumber\\
&&\quad\quad\quad\quad+(h^{\vec{n}}+4\phi^{\vec{n}})i\gamma^{\mu}\partial_{\mu}]\tilde{\psi}
\},
\end{eqnarray}
where $\Psi$ and $\tilde{\psi}$ are background field and fluctuation of the Fermion. Our final result for these two diagrams is
\begin{eqnarray}
\Sigma^R(\ps)&=&-i\frac{\hat{\kappa}^2}{128\pi^2}[\frac{\Lambda^n}{(4\pi)^{n/2}\Gamma(n/2)n(n+2)}]{1 \over \epsilon}
(n-16)p^2\ps\nonumber\\
 &&+i\frac{\hat{\kappa}^2}{64\pi^2}[\frac{\Lambda^{n+2}}{(4\pi)^{n/2}\Gamma(n/2)(n+2)^2}]{1 \over \epsilon} (30-12n)\ps+ \; [finite\;\; part]. \label{result2}
\label{resultf}
\end{eqnarray}
After a Fourier transform to the configuration space, the first term
corresponds to a dimension six operator
$\frac{1}{M^2}\bar{\Psi}\pars\pars\pars\Psi$. The second term
contributes to the renormalization of the fermion field.

\begin{figure}[t]
\begin{center}
\includegraphics[width=16cm,clip=true,keepaspectratio=true]{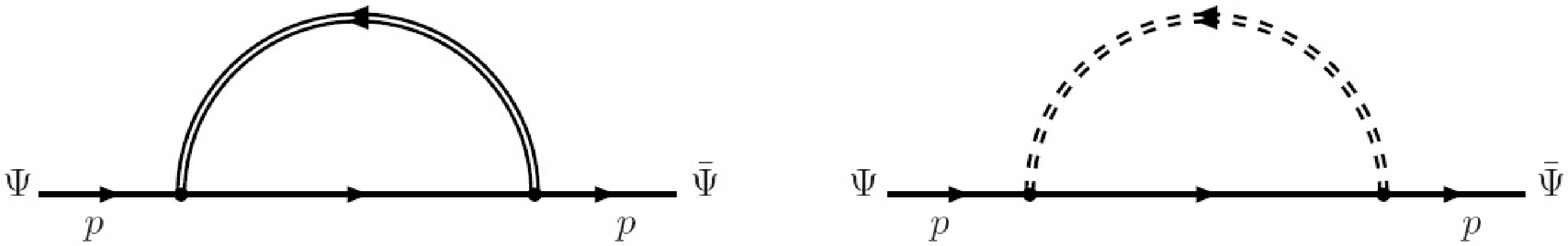}
\caption{\small The rainbow diagrams generating the higher
derivative term of fermion. The internal line, double lines and dash
double lines represent $\tilde{\psi}$, $h^{\vec{n}}_{\mu\nu}$ and
$\phi^{\vec{n}}_{ij}$ fields respectively. The external lines are
background fermion fields.}
\end{center}\label{diagram2}
\end{figure}

In the effective theory, we should add the higher derivative term
permitted by symmetry at the beginning. Thus the modified curved spacetime
Dirac action becomes
\begin{eqnarray}
\int d^4xd^ny [\sqrt{ - |\hat{g}^{(4)}|}\bar{\psi}(i\hat{e}^{\mu}_{\,\alpha}\gamma^{\alpha}\mathscr{D}_{\mu}+i{a_2\over
M^2}\hat{e}^{\mu}_{\,\alpha}\hat{e}^{\nu}_{\,\beta}\hat{e}^{\rho}_{\,\delta}\gamma^{\alpha}\gamma^{\beta}\gamma^{\delta}
\mathscr{D}_{\mu}\mathscr{D}_{\nu}\mathscr{D}_{\rho})\psi\delta^{(n)}(y)],  \label{moddirac}
\end{eqnarray}
where the derivative is defined as $\mathscr{D}_{\mu}=D_{\mu}+\frac{1}{2}\hat{\omega}_{\mu ab}\sigma^{ab}$. The effective action in the
four-dimensional Minkowski spacetime is of the form
\begin{equation}
\int d^4x\, \bar{\psi}(i\Ds+i{a_2\over M^2}\Ds\Ds\Ds)\psi.  \label{diracLW}
\end{equation}
The propagator of the higher derivative theory is
\begin{equation}
{i\over \ps-{a_2\over M^2}p^2\ps}={-i\over {a_2\over
M^2}(p^2-{M^2\over a_2})\ps}, \end{equation} which implies two
freedoms of fermion: One is the massless fermion, the other one is
its Lee-Wick partner with mass ${M\over \sqrt{a_2}}$.

\subsection{Lee-Wick Scalars}
The fourth term in the action (\ref{action}) is the Klein-Gorden theory in the curved spacetime. In terms of the Kaluza-Klein states of the graviton, the
relevant interaction terms can be reduced to
\begin{eqnarray}
&&-{\kappa\over 2}\sum_{\vec{n}}h^{\vec{n}}_{\mu\nu}[2\partial^\mu\tilde{\varphi}\partial^\nu\Phi-\eta^{\mu\nu}
(\partial_{\lambda}\tilde{\varphi}\partial^{\lambda}\Phi-m^2_s\tilde{\varphi}\Phi)]
+\kappa\sum_{\vec{n}}\phi^{\vec{n}}(\partial_{\lambda}\tilde{\varphi}\partial^{\lambda}\Phi-2m^2_s\tilde{\varphi}\Phi).
\end{eqnarray}
Similar to the gauge and fermion field sectors, only the rainbow
diagrams plotted in Fig. 3 contribute to the higher derivative
operator. Summing up these two diagrams, we find the final result
\begin{eqnarray}
\Omega^R(p^2)&=&-i\frac{\hat{\kappa}^2}{16\pi^2}\frac{\Lambda^n}{(4\pi)^{n/2}\Gamma(n/2)}{1 \over \epsilon}
({1\over n+2}p^4+{n+16\over n(n+2)}m_s^2p^2\nonumber\\
 &&+{5n+8\over (n+2)^2}\Lambda^2p^2+{8n-16\over n(n+2)}m_s^4)+ \; [finite\;\; part].
\label{results}
\end{eqnarray}
It is easy to transform the first term to the configuration space.
It is a higher derivative operator ${1\over M^2}(\partial^2\Phi)^2$.
The other three terms work on the renormalizations of the field and
mass of the scalar. Note that when there are not any extra
dimensions, i.e. $n=0$, the higher derivative operator does not
appear.

\begin{figure}[t]
\begin{center}
\includegraphics[width=16cm,clip=true,keepaspectratio=true]{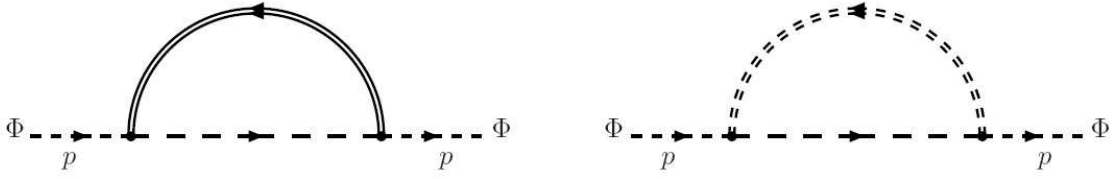}
\caption{\small The rainbow diagrams generating the higher
derivative term of scalar. The internal dash line, double lines and
dash double lines represent $\tilde{\varphi}$,
$h^{\vec{n}}_{\mu\nu}$ and $\phi^{\vec{n}}_{ij}$ fields
respectively. The external dash lines are background scalar fields.}
\end{center}\label{diagram3}
\end{figure}

To write down a one-loop renormalizable theory, we should include the higher derivative operator allowed by symmetry in action (\ref{action}). Then the
modified action has the form
\begin{eqnarray}
{1\over2}\int d^4xd^ny \sqrt{ - |\hat{g}^{(4)}|}[\hat{g}^{\mu \nu}\partial_{\mu}\varphi\partial_{\nu}\varphi-{a_3\over M^2}(\hat{g}^{\mu \nu} {\cal
D}_{\mu}{\cal D}_{\nu}\varphi)^2-m^2_s\varphi^2]\delta^{(n)}(y).
\end{eqnarray}
By an straightforward reduction, the quadratic terms in the flat four-dimensional world can be obtained
\begin{equation}
{1\over2}\int d^4x[\partial_{\mu}\varphi\partial^{\mu}\varphi-{a_3\over M^2}(\partial^2\varphi)^2-m^2_s\varphi^2].
\end{equation}
One can easily find the propagator to be
\begin{equation}
{i\over p^2-{a_3\over M^2}p^4-m^2_s}.
\end{equation}
For ${M\over\sqrt{a_3}}\gg m_s$, it has poles at $p^2\simeq m^2_s$
and at $p^2\simeq M^2/a_3$, indicating the description of two
degrees of freedom: the scalar and its Lee-Wick partner.

\section{conclusions}
\label{sec4} We have studied Maxwell, Dirac and Klein-Gorden
theories in the model of large extra dimensional gravity in which
the gravitons propagate in the (4+n)-dimensional bulk, while the
gauge and matter fields are confined to the four-dimensional world.
The one-loop corrections to the two-point Green's functions of the
gauge, fermion and scalar fields from the exchange of virtual
Kaluza-Klein gravitons have been calculated in the gauge independent
background field method. We show that the gravitational corrections
generate a new type of divergence which corresponds to a higher
derivative operator with mass dimension six for every field.
Besides, the one-loop corrections to the $\beta$-function of the
gauge coupling from a tower of graviton Kaluza-Klein states is found
to be of order $\Lambda^{n+2}$.

In the framework of effective field theories, the higher derivative
operators permitted by the symmetry should be added to Lagrangian at
the beginning so that one can subtract the divergences and
renormalize the theory. We show that the modified one-loop
renormalizable Lagrangian contains a TeV scale Lee-Wick partner of
every gauge and matter field as an extra degree of freedom. Since
gravitons do not carry gauge charges, the same results can be
applied directly to the non-abelian case. Thus the large extra
dimension model of gravity provides a natural mechanism to the
emergence of these exotic particles which were introduced in
\cite{Grinstein} to construct an extension of the Standard Model.

\section*{Acknowledgements}

The research of F.W. is supported in part by the project of Chinese
Ministry of Education (No. 208072). M.Z. is supported in part by the
research fund of National University of Defense Technology.

\end{document}